\documentclass[apj]{emulateapj}
\input epsf.tex
\begin{document}
\title{Constraining Neutrino Cooling using the Hot White Dwarf Luminosity Function in the Globular Cluster 47 Tucanae} 
\author{Brad M. S. Hansen\altaffilmark{1}, Harvey Richer\altaffilmark{2},  Jason Kalirai\altaffilmark{3,4}, Ryan Goldsbury\altaffilmark{2}, Shane Frewen\altaffilmark{1},  
 Jeremy Heyl\altaffilmark{2} }
\altaffiltext{1}{Department of Physics \& Astronomy, University of California Los Angeles, Los Angeles, CA 90095, hansen@astro.ucla.edu}
\altaffiltext{2}{Department of Physics \& Astronomy, University of British Columbia, Vancouver, British Columbia, V6T 1Z1, Canada}
\altaffiltext{3}{Space Telescope Science Institute, 3700 San Martin Drive, Baltimore, Maryland 21218}
\altaffiltext{4}{Center for Astrophysical Sciences, Johns Hopkins University, Baltimore, MD, 21218}


\shortauthors{Hansen et al}
\shorttitle{Neutrino Cooling of Hot White Dwarfs}

\begin{abstract}
We present Hubble Space Telescope observations of the upper part ($T_{\rm eff}> 10^4$ K) of the white dwarf cooling sequence in the globular cluster 47~Tucanae
and measure a luminosity function of hot white dwarfs. Comparison with previous determinations from large scale field surveys indicates
that the previously determined plateau at high effective temperatures is likely a selection effect, as no such feature is seen in this
sample. Comparison with theoretical models suggests that the current estimates of white dwarf neutrino emission (primarily by the plasmon channel) are accurate, and variations are
restricted to no more than a factor of two globally, at $95\%$ confidence. We use these constraints to place limits on various proposed exotic
emission mechanisms, including a non-zero neutrino magnetic moment, formation of axions, and emission of Kaluza-Klein modes into extra dimensions.
\end{abstract}


\section{Introduction}

Despite its modest beginnings in a sheepish proposal by Pauli in his famous letter of 1930\footnote{Reprinted, in translation, in
Brown 1978}, the neutrino has
emerged as an object of great interest to the contemporary particle physics community.
The interpretation of solar and atmospheric neutrino data (Davis, Harmer \& Hoffman 1968; Fukuda et al. 1998) in
terms of neutrino oscillations requires that neutrinos have finite mass and exhibit flavour mixing, which is
one of the few pieces of evidence that currently points to the existence of physical phenomena beyond the standard model of particle physics.
However, the 
 small interaction cross-section of neutrinos means that they are quite difficult to produce and to study in
terrestrial experiments. On the other hand, certain classes of stars produce neutrinos in copious quantities and
neutrino cooling plays an important role in certain aspects of stellar evolution.

Indeed, when
 central
densities $\rho > 10^4 g.cm^{-3}$ and central temperatures are $\sim 10^8$~K, the stellar luminosity in neutrinos
can exceed that in photons (Fowler \& Hoyle 1964; Vila 1966; Lamb \& Van Horn 1975). 
Such conditions are realised during the core evolution of intermediate mass stars, and neutrino cooling regulates
the rate of evolution of stars as they transition from the Asymptotic Giant Branch to the top of the white dwarf
cooling track.
Models for masses $\sim 0.53 M_{\odot}$, an appropriate mass for those white dwarfs forming in old globular clusters today, show that 
for the first $\sim 5 \times 10^6$ years of the white dwarf existence, the rate of cooling is dominated by neutrino emission.
Thus, if we can amass a large enough sample of white dwarfs with ages $<$10~Myr, the study of the resulting hot white dwarf
 luminosity function  can provide a direct test of our understanding of
neutrino emission in astrophysical settings.

Such studies became a possibility with the construction of a luminosity function for hot white dwarfs in the
solar neighbourhood from the Palomar-Green survey (Fleming, Liebert \& Green 1986). The resulting comparison with
 models
for the population evolution of young white dwarfs enabled tests of our understanding of neutrino emission, including
constraints on the possibility of extra cooling from hot cores due to the existence of processes beyond the standard
model, such as axion emission or the existence of a neutrino magnetic moment (Wang 1992; Blinnikov \& Dunina-Barkovskaya 1994).
However, the short evolutionary timescales and relative rarity of hot white dwarfs meant that these constraints were
not improved until the advent of the modern era of large scale sky surveys, such as the Sloan Digital Sky Survey (SDSS -- York et al. 2000) or
SuperCosmos Sky Survey (SSS -- Hambly et al. 2001). This has led to a resurgence of interest in this question, with improved constraints on the
hot white dwarf luminosity function presented by De Gennaro et al. (2008), Krzesinski et al. (2009) and Rowell \& Hambly (2011)
and resulting theoretical analyses (Isern et al. 2008; Miller Bertolami 2014).

However, the construction of a luminosity function from an all sky survey must confront a variety of selection effects,
including a non-uniform selection for spectroscopic follow-up and biases in the assignment of absolute magnitudes due
to distance uncertainties (see Krzesinski et al. 2009 for a discussion). As a result, we describe here the construction
of a hot white dwarf luminosity function from an alternative source, namely the globular cluster 47~Tucanae. This project
is a follow-up to a previous deep observation to constrain the luminosity function of cool white dwarfs (Hansen et al. 2013; Goldsbury et al. 2012)
but with a wider field coverage (albeit to shallower depth) and UV exposures, in order to increase the number of bright white dwarfs detected.
This program has an advantage over all-sky surveys in that all the white dwarfs lie at the same distance and are selected in a
uniform manner, albeit with its own selection effects, to be discussed below. As such, our resulting luminosity function provides
a constraint on hot white dwarf physics that is completely independent of that derived from the stars
in the solar neighbourhood.

We describe below (\S~\ref{Obs}) two sets of observations taken with the Hubble Space Telescope in 47~Tucanae. In \S~\ref{Atmo}
we compare these to a set of white dwarf atmospheric models and describe
resulting construction of the hot white dwarf bolometric luminosity function. We also describe the construction of a set of 
theoretical cooling models
(\S~\ref{Model}) used to explore the possible ranges of physical inputs for comparison to the observations, including
both explorations of the neutrino physics, possible non-standard contributions to the luminosity, and potential sources
of confusion, such as uncertainties in the thickness of the Hydrogen layer on the surface of white dwarfs. Finally, in
\S~\ref{Results} we compare the models and the data to place constraints on the various parameters discussed.

\section{Observational Design and Data Analysis}
\label{Obs}

The observational design of this program (GO-12971; PI H.\ Richer) involves parallel imaging observations 
with {\it HST's} WFC3 and ACS cameras.  The data were collected from Nov 14th 2012 to Sep 20th 2013.  
The 10 orbits were split into 10 visits, each one with an orientation 
offset to ensure a 360 degree mapping of the cluster central region when all data are combined together.  
In this arrangement, the center of 47 Tuc was placed near the corner of all WFC3 observations, so the 
contiguous region from the mosaic extends approximately one WFC3 diagonal field in radial extent.  The 
parallel ACS pointings approximately abut  the WFC3 fields and therefore complete an annulus 
around the cluster center.  Taken together, this arrangement provides high-resolution {\it HST} imaging 
covering the inner 8.5 arcmin of 47~Tuc. 

The data analysis was performed using the prescription and procedures described in Kalirai et~al.\ (2012).  
Within each visit, the observations contain two WFC3 exposures in $F225W$ (380s and 700s) and $F336W$ 
(485s and 720s), and two ACS exposures in $F435W$ (290s and 690s) and $F555W$ (360s and 660s).  For the 
WFC3 observations, we retrieved the \_raw and \_flt frames and corrected the data for CTE losses using 
the WFC3 pixel-based empirical CTE correction software (http://www.stsci.edu/hst/wfc3/tools/cte\_tools).  
For the ACS data, we retrieved the \_flc (i.e., CTE corrected) frames directly 
from the MAST archive.  For each filter on the WFC3 data, we executed a first pass of MultiDrizzle 
(Fruchter \& Hook 2002) to generate geometric distortion corrected images. We then performed astrometry on 
the stars in these images (using the DAOPHOT~II software; Stetson 1987; 1994) and calculated 
transformations that mapped all images on the 10 visits to the same frame of reference.  These offsets and 
rotations were then supplied to MultiDrizzle in a second pass as a ``shift'' file and a stacked (i.e., 
drizzled) image was produced in each filter for the entire 10 visit mosaic.  Given the small overlap 
of the 10 pairs of images, we did not rescale the images or adjust the ``pixfrac'' parameter in 
MultiDrizzle.  As the ACS data contains negligible overlap between the ten fields, each of the 
fields were treated independently to create ten drizzled images from the pairs.

The single drizzled WFC3 image in each filter, and the ten drizzled ACS images in each filter, were subjected 
to PSF-fitted photometry and morphology using DAOPHOT~II.  As described in Kalirai et~al.\ (2012), our 
procedure involved finding all sources that are at least 2.5$\sigma$ above the local sky, selecting 
candidate PSF stars based on their brightness and isolation, generating a spatially variable PSF 
from these stars, and applying the PSF to all sources in the image using ALLSTAR.  The resulting 
photometry from each filter was zero pointed to the VegaMAG system, and the photometry from the two 
filters on each camera was matched into a single catalog.

\subsection{The Core Field}
\label{WFC3}

\begin{figure}
\plotone{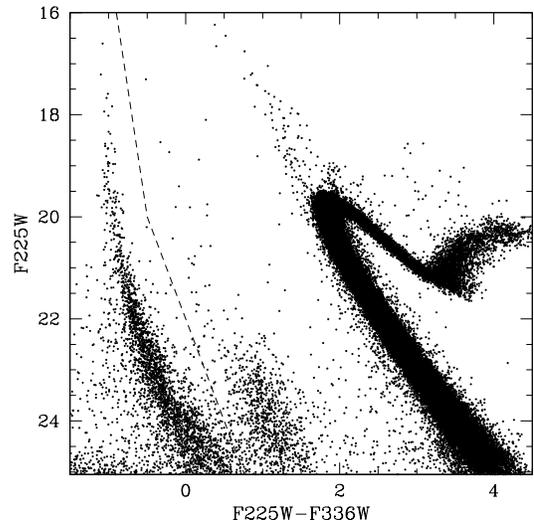}
\figcaption[UB3.ps]{The panel shows the WFC3 UV data on the core field, with the most prominent being
the cluster main sequence and giant branch. A prominent white dwarf cooling sequence is also evident,
and separated from the background SMC population by the dashed curve. The faint population of detections
blueward of the white dwarf sequence are spurious.
\label{UB3}}
\end{figure}

 The CMD of all stars in the WFC3/UVIS drizzled image is shown in Figure~\ref{UB3}, and shows a 
clear white dwarf cooling sequence extending to $F225W \sim 25$. At these short wavelengths, the 
main sequence of the background SMC is less prominent than at other wavelengths (e.g. Hansen et al. 2013; Kalirai
et al. 2013; 
and \S~\ref{Swathe}). The dashed curve shown in Figure~\ref{UB3} is used to identify the 
white dwarf population for the purposes of constructing a luminosity function.

The approximate transition temperature between neutrino-dominated cooling and photon-dominated cooling
is $\sim 25,000$K.
A white dwarf with this temperature  and $\log g=8$, at a distance of 
$\mu_0=13.3$ and a reddening of E(B-V)=0.04 (nominal 47 Tuc parameters), will have $F336W\sim 22.1$. Our luminosity function
contains 216 white dwarfs brighter than this threshold.
Although the exact value of
this transition will vary depending on the distance, extinction and white dwarf parameters, it is clear that our data contains
a substantial population of white dwarfs whose principal cooling mode is via the emission of
neutrinos.

To assess the photometric and astrometric error distribution of the final photometry in our images,
as well as the completeness of the data reductions, we performed artificial star tests.  The artificial
stars are modeled from the stellar PSF and scaled to reproduce the complete luminosity
range of real stars in the drizzled images of each filter.  The input stars were placed along each of
the white dwarf and main sequence of 47~Tuc, and the input positions were set such that no two
artificial stars overlapped each other on the image.  As the mosaicing of our data set results in
deeper observations near the center of the cluster (i.e., exposures from all visits overlap in the core),
we placed the artificial star tests over the complete radial range of our fields.  The procedure to do
this involved placing stars with the same intensity in radial grids.  A total of 25 million artificial
stars were used.  The images with artificial stars were subjected to the photometric routines that were
applied to the actual drizzled images, using identical criteria.  The stars were recovered blindly and
automatically cross-matched to the input starlists containing actual positions and fluxes. 

\subsection{The Outer Fields}
\label{Swathe}

\begin{figure}
\plotone{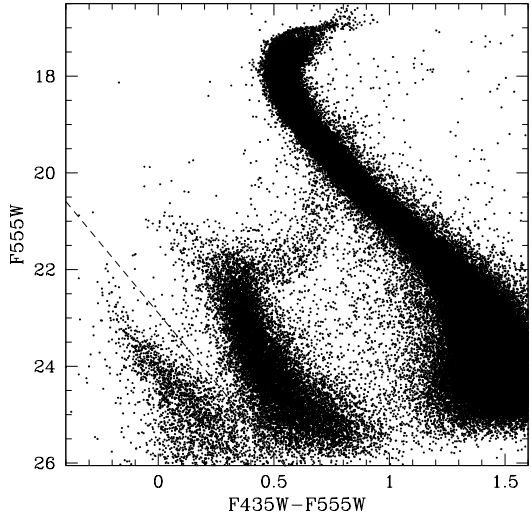}
\figcaption[ACSFinal.ps]{This figure shows the data from the ACS visible CMD compiled
from 9 separate pointings in the outer parts of the cluster.
The three prominent populations are, from the right/red, the cluster
main sequence, the SMC main sequence and the cluster white dwarf population. The dashed line
indicates the CMD cut used to isolate the white dwarf population. We see that this separation becomes
more difficult at fainter magnitudes because the photometric scatter of the white dwarfs and the 
SMC background begin to overlap.
\label{ACScut}}
\end{figure}

Figure~\ref{ACScut} shows the  CMD from the ACS parallel fields.
The presence of the SMC main sequence is much more prominent in these redder filters, and the
photometric scatter is larger because the exposure time per object is $\sim 10\%$ that of the
stars in the core data. As a result, the definition of the white dwarf cooling sequence becomes
more ambiguous for $F555W>24.5$, as the white dwarfs and SMC main sequence start to overlap.
Nevertheless, we can still see a well sampled luminosity function for bright white dwarfs.

The increased field coverage of the ACS data partially compensates for the lower stellar
density relative to the core. The stellar density at $\sim 3$--4 arcminutes (the distance of the parallel fields
from the cluster center and slightly larger than the cluster half-mass radius -- Trager, Djorgovski \& King 1993) is more than an order of magnitude lower than that averaged over the WFC3 field (which encompasses $\sim 2.7$~arcmin), so that the increase in areal coverage by a factor of 10 results in the detection of 89 white dwarfs with $F555W<23.5$ (the magnitude for white dwarfs with $T_{eff}=25,000$K and the same distance and extinction as used for the
WFC3 estimate). 

We follow the same artificial star procedures as in \S~\ref{WFC3} to estimate completeness and photometric scatter. The radial dependance is much less in the ACS
fields, because they lie at several core radii from the cluster, where the density gradient is much smaller across the field.

\section{Fitting Atmospheric Models to the Data}
\label{Atmo}

Our two data sets represent two observations of the same class of stars in multiple bandpasses and spatial
locations in the cluster. The underlying atmospheric models should be the same in both bandpasses, and so
we can hope to constrain the distance and extinction to the cluster based on fitting the position of
the cooling sequence in both colour-magnitude diagrams simultaneously. 

\begin{figure}
\plotone{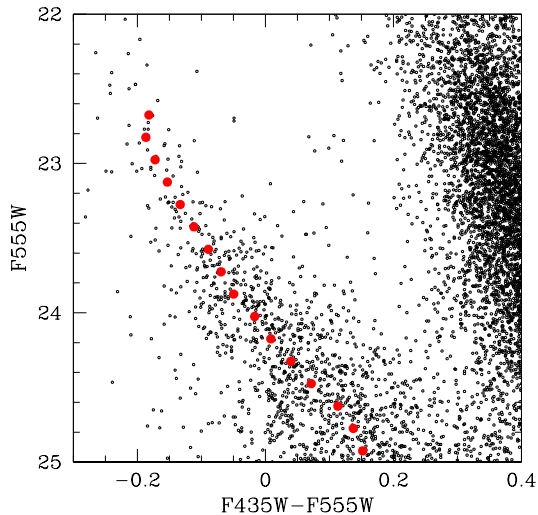}
\figcaption[CMD2.ps]{The small open points represent the observed white dwarf population in the swathe of outer fields
observed with the ACS camera. The large red points show the best fit atmosphere model, for a white dwarf of mass
$0.53 M_{\odot}$, distance modulus $\mu_0=13.21$ and E(B-V)=0.04. The best fit is determined by minimising the
difference between the model colour and the mean observed white dwarf colour at that magnitude.
\label{CMD2}}
\end{figure}

\begin{figure}
\plotone{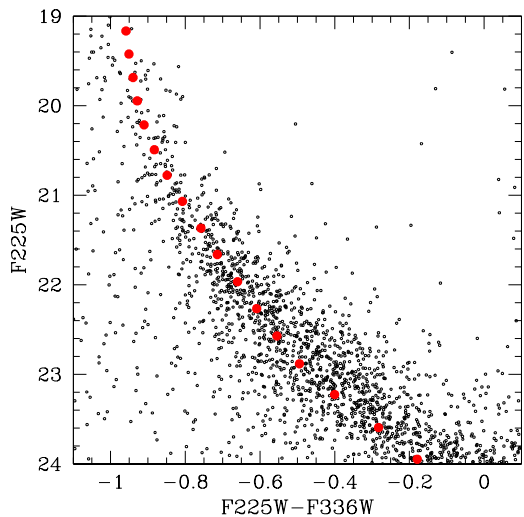}
\figcaption[CMD1.ps]{The small open points represent the observed white dwarf population in the core of 47~Tuc, 
observed with the WFC3 camera. The large red points show the atmosphere model, for a white dwarf of mass
$0.51 M_{\odot}$, distance modulus $\mu_0=13.2$ and E(B-V)=0.02. These are chosen to represent the most plausible
fit consistent with other determinations of distance. We see that the models fit the data well for F336W$<23$ but
are too blue relative to the data for fainter magnitudes. It is this mismatch that drives the best fits to lower
distances.
\label{CMD1}}
\end{figure}

In \S~\ref{Results} we will fit the data to models that include cooling histories, but we can compare the
model and observed colours independently of the cooling history.
 This is because
the position of the cooling sequence in the colour-magnitude diagram depends on the model atmosphere, while
the distribution of stars {\em along} the cooling sequence is determined by the rate of cooling. To constrain
the former alone, we calculate
the mean white dwarf colour as a function of magnitude (in 0.25 magnitude bins) in both the core and swathe fields, and fit an
atmosphere model for a range of assumed white dwarf mass, extinction and distance, using the artificial
star tests to estimate photometric scatter as a function of magnitude. We also considered various values of
the surface Hydrogen layer mass, which can inflate the radius of hot white dwarfs. Figures~\ref{CMD2} and \ref{CMD1} shows the resulting
best fits in both cases.

For the ACS data, the best fit distance is $\mu_0=13.21 \pm 0.16$, for an extinction $E(B-V)=0.04 \pm 0.02$. The best
fits are obtained for masses $\sim 0.51-0.53 M_{\odot}$, and Hydrogen envelope  mass fractions $q_H = 10^{-4}$. The error bar on this determination
is larger than our previous measurements in Woodley et al (2012) and Hansen et al. (2013) because the more limited
magnitude range of the data allows for more covariance between distance, extinction and mass than in those analyses. The
value here is slightly lower than in those cases $13.36 \pm 0.08$ and $13.32 \pm 0.09$ respectively), but still consistent within the error bars.
The constraint will be improved in \S~\ref{Results} when we use the full information in the cooling models.

The best fit distance for the core data is $13.08 \pm 0.08$, again favouring white dwarf masses $\sim 0.51-0.53 M_{\odot}$ but 
now the best fits are obtained for low extinctions and large Hydrogen layer masses. Indeed, inspection of the comparison between
models and data in Figure~\ref{CMD1} suggests that the atmosphere models are not a good representation of the data in these passbands, as they are 
too blue at magnitudes $F336W>23$. Furthermore, the fits favour low extinction because the colour asymptotes to a value at bright
magnitudes that is a poor match to the observed values, which drags the fitted extinction down and increases the mismatch at the
faint end. The modelling of the core white dwarf sequence is further complicated by the effects of spatially varying completeness
and the possibility of radial diffusion of stars relative to dynamical relaxation. However, neither of these effects can explain
the discrepancy here, because the best-fit colour is determined at each magnitude independently, and does not depend on the relative
numbers as a function of magnitude.

\subsubsection{Causes of the Discrepancy}

A variety of tests were performed to identify the source of the discrepancy between the data and the available atmosphere models at the faint end of the UV white dwarf cooling sequence.  These included re-drizzling smaller sections of the image mosaic that excluded the dense core of 47~Tuc and performing PSF photometry on these regions, adjusting the sky subtraction algorithm to correctly subtract the diffuse light between different frames, adjusting the annulus over which the sky is measured in the photometry to test for red leaks that were not being accounted for, and exploring a range of apertures in a separate trial of aperture photometry analysis of the non crowded regions.  We also performed a completely independent analysis of the data without the drizzling algorithm and with a different PSF-fitting photometric algorithm (J.\ Anderson, priv comm.).  All of these tests resulted in colors for the faint white dwarfs consistent with our baseline reduction.

Subsets of our core field mosaic overlap observations in other WFC3 UV filters, including $F275W$ (GO-12311; PI.\ G.\ Piotto) and $F390W$ (GO-11664; PI.\ T.\ Brown).  We retrieved all of these data from MAST and subjected them to the same drizzling and PSF-fitting operations as our baseline data set, and matched the positions of sources across all four filters.  While these data do not have the same depth and spatial coverage of our new observations, they hint that the color offset with respect to the models
may also appear in other color combinations that span a wavelength $\sim 300$nm (e.g., a possible offset is also seen in a $F275W$ vs $F390W$ comparison). 

The source of this mismatch remains a mystery, and could be the consequence of one or more independent reasons.  For example, the images in the redder filters like $F336W$ and $F390W$ have a very high level of crowding as we approach the cluster center, and therefore the completeness is $<$75\% even at 2 core radii.  So, if there is a systematic issue with the photometry that is not being accounted for by the artificial star tests, then incompleteness could be the cause. Alternatively, it is possible that the observed colours
are correct
 and rather the current generation of white dwarf atmosphere models lack an important source of opacity at UV wavelengths.

\subsection{The Bolometric Luminosity Function}
\label{Bolo}

The data presented here represent a new opportunity to construct a luminosity function for hot white dwarfs. This is
particularly difficult in the field because the short cooling times imply that hot white dwarfs are relatively rare,
requiring large scale surveys to characterise them. Observations in a globular cluster can isolate a well defined
sample and better illuminate features of the luminosity function, at the price of foregoing some of the more detailed
atmospheric characterisation possible from spectra of the closer field stars. In \S~\ref{Results} we will provide a
detailed model fit to the data, but for the purposes of comparison to prior studies, it is also worthwhile to present
the data in the same form as in other studies.

The construction of an absolute bolometric magnitude requires the adoption of a cluster distance and reddening. Fortunately,
the sparseness of the data in wide field surveys means we can compare results in relatively large (1 magnitude) bins, which
renders the uncertainties in the discussion above (of the order of 0.1 magnitudes) irrelevant.
We will adopt $\mu_0=13.3$ and $E(B-V)=0.04$, consistent with our determination above and our prior analyses of this cluster.
 In later sections, we
will allow these quantities to vary in the fitting procedure, but we keep them fixed in this case because the uncertainties
are small relative to our bin size.

Our atmosphere model colours are based on
pure Hydrogen atmosphere models from Tremblay, Bergeron \& Gianninas (2011), as a function of $T_{eff}$ and 
surface gravity. In the case of the WFC3 field, the F336W data are the deepest (photometric errors $<0.1$ magnitude at 
a fainter position along the cooling sequence) and, in the case of the
ACS fields, the F555W data are deepest. Thus, we convert the F336W and F555W magnitudes to bolometric
magnitudes using the above models.  The assumption of a pure Hydrogen atmosphere model for the hottest
white dwarfs is questionable because it takes a finite time for the heavy elements to sediment out and
for a star to take on the traditional onion-skin structure assumed for white dwarfs. Fortunately, the spectroscopic
classifications of the SDSS hot white dwarfs from Krzesinski et al. (2009) clearly identify this transition,
showing that DA white dwarfs appear for bolometric magnitudes $M_{bol} > 1.5$. Thus, for magnitudes brighter
than this, our bolometric magnitudes may be inaccurate, but the bulk of our sample is fainter than this.

To construct the bolometric luminosity function in the outer fields, we must account for the effects of
photometric incompleteness.
 We first count and bin observed stars as a function
of F555W to obtain a raw number count. In order to understand the incompleteness, we perform a monte carlo
simulation in which we draw white dwarfs of the appropriate magnitude 
 and then quantify their expected detectability using the results of the artificial star
tests. This is sampled until the number of detections equals the observed raw number, and the number of drawn
input stars required to meet this target
 yields an estimate of the true underlying number of white dwarfs of this magnitude in our field.

In the case of the luminosity function in the core, we  must also
 account for the fact that the density of sources varies across the field, and this leads to a
covariance between incompleteness and location in the field.
Our artificial star tests determine incompleteness as a function of radius,
 and so we need to account for the radial profile of the underlying sources as well.
 Fortunately, 47 Tuc has been a subject of close study for many years, and the
radial profile of stars is known to be well described by a King model with $W_0=8.6$ (McLaughlin et al. 2006).
Our core field extends out to $2.7'$, and so we are able to measure a core radius for MS stars directly
from our data, yielding $r_0 = 29.7'' \pm 0.4''$. This is within $2\sigma$ of that measured by
Goldsbury, Heyl \& Richer (2014) from surface density profiles.

The core radii
 of white dwarfs and main sequence stars may not be the
same, however, because of the mass loss associated with the latter stages of stellar evolution. The energy
equipartition in such a dynamically relaxed system implies that lower mass stars such as white
dwarfs ($\sim 0.5 M_{\odot}$) should have a larger velocity dispersion and therefore larger radius
than more massive main sequence turnoff stars ($\sim 0.8 M_{\odot}$). Thus, our modelling needs to
consider a range of possible core radii. Furthermore, there is a distinct possibility that the 
core radius may be a function of white dwarf luminosity. If the core relaxation time is longer than
the timescale over which a star loses mass, then the hottest white dwarfs may still exhibit the
core radius of their more massive progenitor population. This results in an intimate connection
between the details of stellar evolution and the number counts of hot white dwarfs. If much of the
mass is lost on the Asymptotic Giant Branch, then the hottest white dwarfs may be over-represented
in the core relative to their cooler brethren, as they will still exhibit a smaller velocity dispersion.
Alternatively, if much of the mass is lost on the red giant branch, then we expect a smaller effect
of this sort, because the stars will have spent $\sim 10^8$ years on the Horizontal branch with a
mass close to that of a white dwarf, and so will have adjusted their velocity dispersion appropriately
before ever becoming a white dwarf.
Thus, we wish to examine a range of core radii ranging from one similar to the MS stars, up to 
 $\sim 1.27 \times 29.7'' \sim 38''$, based on a white dwarf mass
of $0.53 M_{\odot}$ and an MSTO mass $\sim 0.86 M_{\odot}$ (e.g. Thompson et al. 2010). 

The construction of our resulting empirical luminosity function proceeds as follows when using the F336W data,
mapped onto bolometric magnitudes using the atmospheric models.
In order to keep our results less sensitive to the modelling, we
 restrict our attention to radii $> 30''$ from the core,
in order to avoid uncertainties associated with the aforementioned dynamical adjustment. On these scales,
the two-body relaxation time locally is $ > 10^8$ years, and therefore should not affect the luminosity
function at magnitudes appropriate to neutrino cooling\footnote{The potential effects of dynamical
relaxation are dealt with in more
detail in  Heyl et al 2015.}. We count and bin observed stars as a function
of F336W to obtain a raw number count. In order to understand the incompleteness, we perform a monte carlo
simulation in which we draw white dwarfs of the appropriate magnitude from the underlying King model, to
form a radial profile and then quantify their expected detectability using the results of the artificial star
tests. This is sampled until the number of detections equals the observed raw number, and the number of drawn
input stars required to meet this target 
 yields an estimate of the true underlying number of white dwarfs of this magnitude in our field.
The error bars are calculated by repeating the procedure for a range of potential core radii from 20''--40''
(to account for uncertainties in the cluster profile) and also by accounting for the Poisson fluctuations
in the raw counts. In this manner, we define an empirical bolometric white dwarf luminosity function in
 the core of 47~Tucanae while accounting for photometric scatter and uncertainties in the mass profile.

\begin{figure}
\plotone{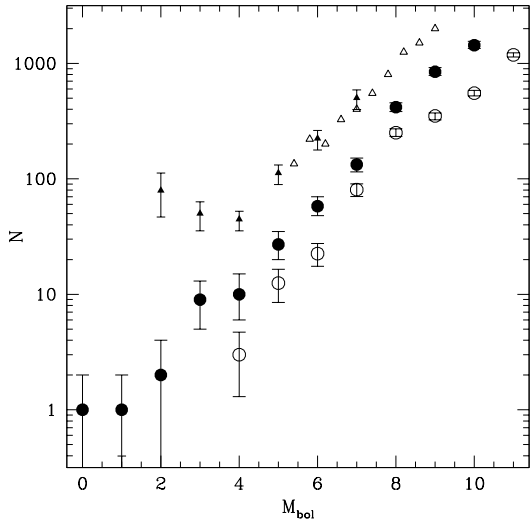}
\figcaption[NewBol.ps]{The filled circles indicate the bolometric function measured for the inner field using the WFC3 data,
and the open circles indicate the equivalent measured using the ACS data in the outer fields. The similarity in the
number counts is somewhat coincidental, resulting from the competing effects of lower stellar density but greater
field coverage in the ACS data. Also shown, as triangles, are the equivalent luminosity functions measured for the disk
by SDSS survey, as presented by De Genarro et al. (2008) -- open symbols -- and Krzesinski et al. (2009) -- filled symbols.
These densities have each been multiplied by an overall normalisation factor to convert them to a number on this plot so that
one can visually compare the resulting slopes. 
\label{NewBol}}
\end{figure}

 The resulting luminosity functions from the two data sets are given in Table~\ref{BolLF}.
Figure~\ref{NewBol} also compares these two independent determinations to the luminosity functions derived
for the solar neighbourhood from the SDSS by De Gennaro et al. (2008) and Krzesinski et al. (2009). In those cases, a correction for the number density of stars has
to be made, so we have allowed for an overall normalisation shift and have set the normalisation so that the
disk and cluster LF do not overlap exactly, to enable ease of visual comparison.
The Krzesinki et al. (2009) luminosity function is the only one that goes to bright enough magnitudes to provide a 
meaningful comparison with our data set. The precise number counts at the bright end of that data are somewhat uncertain because
of several completeness effects related to the spectroscopic selection, as described in Krzesinski et al. and in
Eisenstein et al. (2006). However, the ability to get spectroscopy of their stars allows them to classify stars 
according to spectral type,  for which our stars are too faint.
 As noted above, genuine DA white dwarfs
only appear at $M_{bol} \geq 1.5$ in the SDSS sample, as hotter white dwarfs have not yet had sufficient time for gravitational settling
to occur, and so our bolometric corrections are likely to be inaccurate for $M_{bol}<1$. The slopes of the disk and cluster luminosity functions compare well for $M_{bol}>5$ though, with the exception
of a possibly larger flatening observed in the ACS data for $M_{bol}>8$. At these magnitudes, the scatter in the white dwarf sequence and
the SMC main sequence begins to overlap, and the lower counts are possibly the result of white dwarfs being lost due to confusion with
SMC main sequence stars.

At the bright end, our luminosity functions are consistent with a relatively smooth extrapolation
of the slope at the fainter end towards higher luminosities. This contradicts the expectations
based on the SDSS luminosity function. If we normalise the Krzesinski et al. LF to ours at
$M_{bol}=4$, the SDSS data would predict 10--19 stars (depending on which SDSS LF binning is used)
 in the magnitude bin centered on $M_{bol}=2$, whereas we count only two stars in this bin, a 
discrepancy of approximately 3~$\sigma$.
Torres et al. (2014) have commented
 on the SDSS `plateau' and found that it could not be reproduced with standard models. 
 Torres et al examined a variety of possible explanations and  concluded that the upturn cannot be the result of a feature in
the Galactic inputs such as star formation rate variations. They further concluded that errors in 
the model colours, uncertain white dwarf cooling tracks or statistical fluctuations were unlikely to
be the cause. Their preferred solution was that the observed flatening was the result of erroneous mass determinations
that led to some white dwarfs being assigned unrealistically low masses and consequently larger distances and
larger inferred luminosities. All of our white dwarfs are at the same distance and have the same mass, since
they follow a single cooling track, so that this selection effect should not influence this luminosity function. 
The uncertainty in the bolometric corrections for the core data should also not affect this conclusion, because
the bins are quite large, and any white dwarfs observed would be counted. In summary, the lack of any plateau in
our data confirms the conclusion of Torres et al. that this is likely the result of systematic error rather than
a physical effect. This is consistent with the behaviour of the proper motion selected LF of Rowell \& Hambly (2011)
at the bright end (shown in Figure~3 of Rowell 2013), which also drops monotonically except for a spike at the
brightest magnitudes which may represent the presence of non-DA white dwarfs.

\begin{deluxetable}{lllll}
\tablecolumns{5}
\tablewidth{0pc}
\tablecaption{Bolometric Luminosity Functions for 47 Tucanae. 
\label{BolLF}}
 \tablehead{ & \multicolumn{2}{c}{F336W} & \multicolumn{2}{c}{F555W} \\
 \colhead{$M_{bol}$} & \colhead{N} & \colhead{N'} & \colhead{N} & \colhead{N'} 
}
\startdata
  0 &  1 & 1$\pm 1$      &   0 & \\
  1 &  1 &  1$\pm 1$     &   0 &      \\ 
  2 &  2 & 2$^{+2}_{-2}$ &   0&  \\
  3 &  7 & 9$^{+4}_{-4}$ &   0&  \\
  4 &  8& 10$^{+5}_{-4}$ &   3& 3$^{+2}_{-2}$\\
  5 &  20& 27$^{+8}_{-7}$ &  12& 13$^{+4}_{-4}$\\
  6 &  41& 58$^{+12}_{-10}$ &  21& 23$^{+5}_{-5}$\\
  7 &  89& 133$^{+18}_{-18}$ &  72& 81$^{+10}_{-10}$ \\
  8 &  269& 417$^{+39}_{-34}$ &   214& 251$^{+20}_{-18}$\\
  9 &  513& 845$^{+74}_{-60}$ &   279& 350$^{+22}_{-26}$\\
  10 & 809 & 1436$^{+118}_{-91}$ & 405 & 551$^{+33}_{-29}$ 
\enddata
\tablecomments{Under each bandpass we
list the raw counts ($N$) in each bolometric magnitude bin and then the counts corrected
for selection biases ($N'$) as described in the text. The error bars incorporate the effects
of photometric scatter, incompleteness and the radial density profile, and are thus moderately
larger than the nominal Poisson error.  Note that the bolometric corrections for the brightest
two bins ($M_{bol}<1.5$) may be uncertain as these may be too young to be pure H atmospheres.}
\end{deluxetable}

\section{Models}
\label{Model}

In order to evaluate the role of various physical inputs into the cooling of white dwarfs, we need to
match our observed luminosity function to models for the rate of cooling. We have established an extensive
framework for such modelling in previous papers on the cooling functions of cluster white dwarfs, most
recently Hansen et al. (2007) and Hansen et al. (2013). However, our previous work has been devoted primarily
to the coolest white dwarfs and the focus on hot white dwarfs in this project necessitates the investigation
of several relevant factors with greater care and detail than in our previous work.

\begin{figure}
\plotone{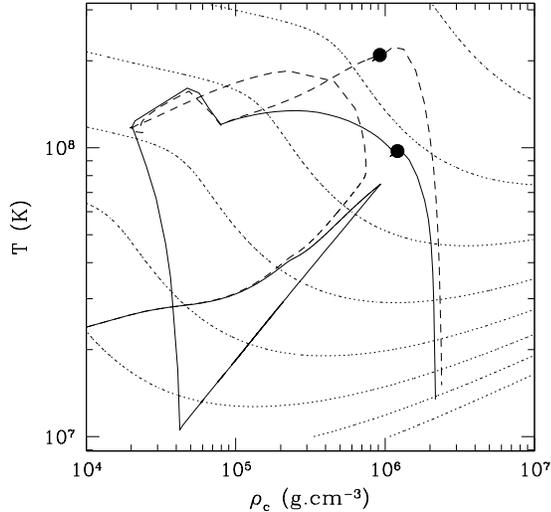}
\figcaption[MESA.ps]{The solid curve indicates the evolution of the central density and temperature of a 
0.9$M_{\odot}$ star with standard input physics, while the dashed line indicates the equivalent but for
a star without neutrino cooling. The two large solid points indicate the location at which each star
passes through an effective temperature of $10^5$K -- an approximate start of the white dwarf cooling
sequence. The dotted contours indicate lines of constant neutrino emissivity, increasing by two orders
of magnitude with each contour as one moves from the lower left to the upper right.
\label{MESA}}
\end{figure}

Our default white dwarf cooling models are based on the code of Hansen (1999). For the
purposes of this work, the most important part is the implementation of neutrino cooling, which 
is taken from the work of Itoh et al. (1996). The focus on young, hot, white dwarfs also requires
greater attention to the starting conditions of stars that begin
to descend the white dwarf cooling sequence. 
 In order to establish a baseline starting model,
we have used the MESA models (Paxton et al. 2011)
to construct full evolutionary models of stars with initial masses $\sim 0.9 M_{\odot}$ and metallicity 0.004.
Figure~\ref{MESA} shows the evolution of the central density and temperature for two stellar models of
initial mass $0.9 M_{\odot}$ and metallicity 0.004, both
with (solid curve) and without (dashed curve) neutrino emission. The dotted contours  indicate loci
of constant neutrino emissivity. The two solid points indicate the location at which the star reaches an
effective temperature of $10^5$~K, an approximate location for the start of the white dwarf cooling
sequence. These curves indicate that a self-consistent treatment of neutrino cooling in the stellar
evolution implies differences of a factor of two in the initial white dwarf central temperature. They
also indicate that the dominant neutrino emission process of relevance here is plasmon emission. The
change in slope of the dotted contours corresponds to the transition from emission due to photoneutrino 
production (left of the diagram) to plasmon decay (lower right). The white dwarfs are well into
this latter regime.

 We use these models as the reference model and consider variations in the input parameters
in our cooling models to investigate the sensitivity to uncertain aspects of the physics.

 The most
obvious parameter of relevance is the total white dwarf mass. We consider a range of masses $\sim 0.5$--0.55$M_{\odot}$.
This is the mass range expected on theoretical grounds (e.g. Fusi-Pecci \& Renzini 1976) and
observed directly from the pressure broadening of spectra of hot cluster white dwarfs (Moehler et al. 2004; Kalirai et al. 2009). 
Modelling of the full white dwarf cooling sequence in this cluster and others (Hansen et al. 2007; 2013)
also agrees with this estimate. In that case, the mass can be constrained directly because the observations contain mass information due
to the fact that the
onset of crystallisation is mass-dependant and so the cooling curves for different masses are not
scaled versions of one another. The sample of white dwarfs observed here is too hot to have undergone crystallisation
and so we restrict our analysis to the plausible mass range above identified by other means. 

 In addition, although we start with a reference model, there is also some uncertainty in the establishment
of this baseline.
 The chemical composition of the white dwarf core is  affected by prior evolution as the temperature
history of the core, as well as uncertainties in the cross-section of some nuclear reactions, will affect the relative abundances of Carbon and Oxygen and their profile through the core.
We have studied this in detail using a range of stellar models, as detailed in the supplementary information
of Hansen et al. (2013). These models also account for possible variations in the initial Helium abundance of the star
and uncertainties related to possible mixing of compositional gradients by Ledoux convection. Incorporating these uncertainties produces
 a range in central Oxygen mass fraction of 0.64--0.86. The core composition
proves to be much less important for the hot white dwarfs discussed here than for the cooler white dwarfs
undergoing crystallisation, because the neutrino emission is quite insensitive to the 
relative proportions of Carbon and Oxygen in the core.
 Furthermore, although the MESA models provide an initial temperature profile, with a central temperature
$\sim 10^8$K with neutrino cooling and $\sim 2 \times 10^8$K without, we also allow for a variation
of 10\% in starting temperature to account for uncertainties in prior evolutionary stages.

We have also examined a range of Helium and Hydrogen layer masses. These lie at sufficiently
low densities that they will not have an effect on the
neutrino emission, but can contribute to the regulation of thermal diffusion of energy through
the white dwarf and can therefore potentially affect the location of the transition from neutrino-dominated
to photon-dominated cooling. Hydrogen layers can have two additional effects. The lower mean molecular
weight of Hydrogen means that the radius of the white dwarf can be non-negligibly affected by the thick
hydrogen layer, at least for hot white dwarfs. This can change the location of the cooling sequence in
the colour-magnitude diagram and therefore affect the fit of the data. 
Furthermore, sufficiently thick
Hydrogen layers can result in a contribution to the luminosity from p-p nuclear burning, which can also
change the cooling curve. 

\begin{figure}
\plotone{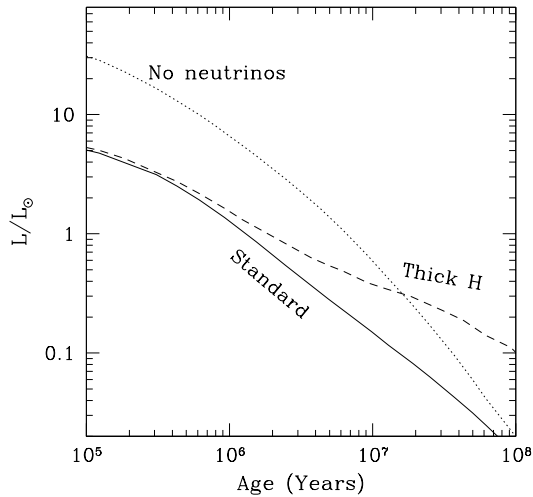}
\figcaption[Comp3.ps]{The solid curve shows the photon luminosity of a 0.53$M_{\odot}$
white dwarf with standard neutrino cooling. The corresponding model with no neutrino luminosity is
given by the dotted curve. The dashed curve represents the same model except that we have
increased the initial H layer mass.
\label{Comp}}
\end{figure}

For young white dwarfs, it is also possible that the H and He layers have not completely
gravitationally separated, which will change the opacity of the atmosphere and affect the
cooling. This occurs near the surface of the white dwarf and will therefore have a 
negligible effect on the neutrino cooling, but it may affect the location where photon
cooling dominates over neutrinos. We note, however, that we examine a large range of
Hydrogen and Helium layer masses, and that the Helium layer masses are an order of magnitude
or more larger than the Hydrogen layer masses. Thus, a surface layer of Hydrogen diluted 
with Helium can be considered the limiting case of a thin Hydrogen atmosphere. We have 
considered H layer mass fractions as small as $q_H=10^{-6}$ and find that our results
are very insensitive to values of $q_H<10^{-4}$. Thus, we believe that uncertainties in
the separation of H and He do not affect our results.

\subsection{Binary White Dwarfs}

The WFC3 data are taken in the cluster core and so any effects of binarity are likely to be
increased due to mass segregation. However, the hot white dwarf sequence observed here covers
a very short-lived phase of stellar evolution and so the observation of two hot white dwarfs
in the same binary would require that the original binary pair had main sequence masses within
1\% of each other. If the mass ratio is determined by randomly sampling the Salpeter mass function,
this occurs in $\sim 10^{-3}$ of pairs containing one main sequence turnoff star. Thus, even for
relatively high global binary fractions, we do not expect a substantial contribution of binaries
to this luminosity function. 

\subsection{Non-standard neutrino physics}

There are many proposals for how modifications to the properties of the neutrino could point
the way to physics beyond that contained in the standard model. One of the most prominent is
the proposal that the neutrino may possess a non-zero magnetic dipole moment. Such a modification
also has an effect on the plasmon neutrino cooling rate and is thus amenable to constraint in
hot white dwarfs (e.g. Wang 1992, Blinnikov \& Dunina-Barkovskaya 1994, Haft, Raffelt \& Weiss 1994).

We investigate the resulting extra cooling by considering the inclusion of an amplification
factor for the  plasmon neutrino emission from Haft et al. (1994)
\begin{equation}
\frac{\epsilon_{total}}{\epsilon_{standard}} = 1 + 0.318\, Q \left( \frac{\mu_{\nu}}{10^{-12} \mu_B}\right)^2
\left( \frac{\bar{h} \omega_p}{10 keV} \right)^{-2} \label{Eps_q}
\end{equation}
where $\mu_{\nu}$ is expressed in terms of the Bohr magneton $\mu_B$ and $\omega_p$ is
the plasma frequency, given by 
\begin{equation}
\bar{h}\omega_p = 0.021 keV \rho^{1/2} \left[ 1 + \left( \rho/1.96 \times 10^6 g/cm^{-3} \right)^{2/3} \right]^{-1/4}.
\end{equation}
 $Q$ is a function that depends on the underlying plasma properties but is $\sim 1$ to within 10\% over the
density and temperature range appropriate to the centers of white dwarfs.

\subsection{Axion Cooling}

Hot white dwarfs are a favourite astrophysical constraint on hypotheses of exotic particles, because
such particles can cause extra cooling if they couple to nuclei, electrons or photons.
One commonly proposed extra cooling mechanism is emission of axions. The axion is a hypothetical
particle (Weinberg 1978; Wilczek 1978) associated with the U(1) symmetry proposed by Peccei \& Quinn (1977)
as a solution to the strong CP problem of particle physics. If such axions couple to electrons they
can result in an increase in the luminosity of hot white dwarfs.

We can treat this by including an extra cooling contribution in the white dwarf models, of the form
\begin{equation}
\epsilon_{ax} = 1.08 \times 10^{23} {\rm ergs\, g^{-1} s^{-1}} F(\rho,T) \frac{g_{ae}^2}{4 \pi} \frac{Z^2}{A} T_7^4, \label{Eps_ax}
\end{equation}
where $T_7$ is the temperature measured in units of $10^7$K, and $F$ is the correction for ionic correlation effects 
from Nakagawa, Kohyama \& Itoh (1987). The factor $Z^2/A$ ranges from 3.6 to 3.8, depending on the core composition.
The axion-electron coupling $g_{ae}$ is related to the mass of the axion $m_{ax}$ by $g_{ae} = 2.8 \times 10^{-14} m_{ax}/1meV$
(up to an unknown angle $\beta$).

\subsection{Radiation into Extra Dimensions}
\label{KK}

Some models of gravitation invoke a version of string theory
that includes additional dimensions that are compactified on dimensions
large enough to have experimental consequences (Arkani-Hamed, Dimopolos \& Dvali 1998).
In such models, the coupling between photons, electrons or nucleons to gravity
can result in the emission of Kaluza-Klein gravitons into these extra dimensions, and will
manifest themselves as an extra source of cooling for hot stars. Barger et al (1999)
present expressions for the emissivity for a variety of processes and place
constraints on the mass scale $M_s$ corresponding to extra dimensions by using astrophysical
constraints from the Sun, red giants, and supernova 1987A. Biesiada \& Malec (2002)
place additional constraints on this mass scale using the pulsational stability of the 
pulsating white G117-B15A, and the fact that electron-graviton bremstrahlung would
accelerate the cooling.

We incorporate the emissivities of Barger et al for three processes, namely photon-photon
annihilation, Gravi-Compton-Primakoff scattering, and gravibremstrahlung. These are
\begin{eqnarray}
\epsilon_{\gamma \gamma} & = & 5.1 \times 10^{-9} {\rm ergs\, g^{-1} s^{-1}} T_7^{9} 
\rho_6^{-1} \left( \frac{M_s c^2} {1 TeV} \right)^{-4} \\
\epsilon_{GCP} & = & 4.5 \times 10^{-6}  {\rm ergs\, g^{-1} s^{-1}} T_7^{7}
 \left( \frac{M_s c^2} {1 TeV} \right)^{-4} \\
\epsilon_{GB} & = & 5.8 \times 10^{-3} {\rm ergs\, g^{-1} s^{-1}}  \bar{Z}_7^2 T_7^{3}
 \left( \frac{M_s c^2} {1 TeV} \right)^{-4},
\end{eqnarray}
where $\rho_6$ is the density in units of $10^6 {\rm g.cm^{-3}}$, $T_7$ is the 
temperature in units of $10^7$ K and $\bar{Z}_7$ is the mean ion charge relative to Nitrogen 
(the value for our model mixtures of Carbon and Oxygen range  from 7.2 to 7.8).
  We have assumed
two macroscopic extra dimensions\footnote{One extra dimension is excluded experimentally by confirmation
of Newtons gravitational force law  
and three dimensions produce rates too low to be tested here.}. 
These rates are calculated for non-degenerate material, although Barger et al. claim that corrections for
degeneracy should be of order unity. In the case of the red giant cores considered in that paper, such an
approximation may be acceptable, but it is likely to be an overestimate under the much more degenerate conditions
appropriate to a white dwarf (a fact that does not appear to have been acknowledged by Biesiada \& Malec).
The latter two processes above depend on electron scattering and therefore the rates should be considerably depressed
in a white dwarf. Although a full rederivation of the rates is beyond this paper, we also test modified expressions
for $\epsilon_{GCP}$ and $\epsilon_{GB}$ in which the rates are scaled by a factor $k T/E_F$, where $E_F$ is the
Fermi energy of the electrons. This limits the fraction of the electrons that participate to only those within $k T$
of the Fermi surface, and which therefore are energetically able to scatter to a state that is not excluded by the
Pauli principle.  The resulting expressions are
\begin{eqnarray}
 \epsilon_{GCP} & = & 2.3 \times 10^{-8} {\rm ergs\, g^{-1} s^{-1}} T_7^{8}
\rho_6^{-2/3} \left( \frac{M_s c^2} {1 TeV} \right)^{-4} \\
 \epsilon_{GB} & = & 3 \times 10^{-5} {\rm ergs\, g^{-1} s^{-1}} \bar{Z}_7^2 T_7^{4}
\rho_6^{-2/3} \left( \frac{M_s c^2} {1 TeV} \right)^{-4}. 
\end{eqnarray}

\subsection{Monte Carlo Model}
\label{Monte}

Using the cooling curves constructed in this fashion, we simulate artificial
white dwarf populations assuming that the birth rate of white dwarfs has
been constant over the last $10^9$~years (since our comparison data set
comprises ages $\sim 10^8$ years at the faint end). For each white dwarf,
the model provides a unique luminosity and effective temperature, which
is converted into a magnitude and colour based on a chosen value of 
distance and extinction. We then use the results of our radial density profiles and photometric
scattering matrix to probabilistically assign output magnitudes and
colours given the original input value, including the nondetection of stars. 

\begin{figure}
\plotone{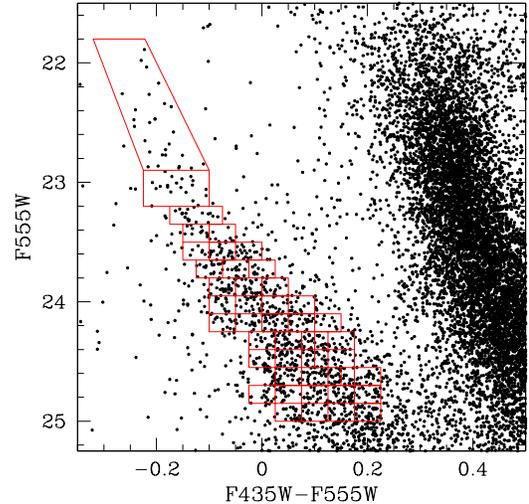}
\figcaption[ACSgrid.ps]{The points show the observed white dwarf cooling sequence
in the ACS bandpass, while the red boxes indicate the grid used to bin the data and
models for the purposes of a quantitative comparison.
\label{ACSgrid}}
\end{figure}

The resulting model populations are then binned in colour and magnitude on the
grid shown in Figure~\ref{ACSgrid}, and compared to the data binned in the same
manner. Given the problematic colours for the core data, we restrict this
analysis to the swathe of outer cluster fields observed with the ACS camera.
The grid is chosen to provide sufficient colour resolution to constrain the
distance and extinction, while still maintaining enough points to ensure stability
of the $\chi^2$ fitting procedure. The brightest bin covers 1 magnitude in range
because the rapid cooling at these temperatures makes the counts insensitive to
the cooling and more sensitive to initial conditions.

We restrict our attention to the stars with F555W$<25.15$. Our limit is set by
the increasing risk of confusion due to photometric scatter from the SMC main
sequence in the background. We also account for this in our reddest bins by
modelling the colour spread of the SMC population as a function of magnitude.
We extrapolate this model bluewards with a maxwellian function, using the fitted
colour dispersion, in order to anticipate the pollution of the white dwarf sequence
with scattered main sequence stars, and include these in the model.

\section{Results}
\label{Results}

 We compare our data with models calculated as described in \S~\ref{Model}, using the
reference model from MESA and considering variations in initial temperature and core chemical
composition.
For each core model, the comparison with the data also requires a choice of distance modulus
and extinction. These, combined with the parameters of white dwarf mass, and Helium and
Hydrogen surface mass fractions, as well as an overall global normalisation, imply a total of six parameters to be varied in the
comparison of a particular model with the data.
 In all cases, we allow the distance modulus to vary from 13.0 to 13.5 and
the reddening to vary from E(B-V)=0.0 to 0.08. White dwarf masses are varied from $0.5 M_{\odot}$ to $0.6 M_{\odot}$,
while Helium and Hydrogen mass fractions are varied from $q_{He} = 0.01$--0.04 and $q_H = 1 \times 10^{-6}$--$ 6 \times 10^{-4}$.

\begin{figure}
\plotone{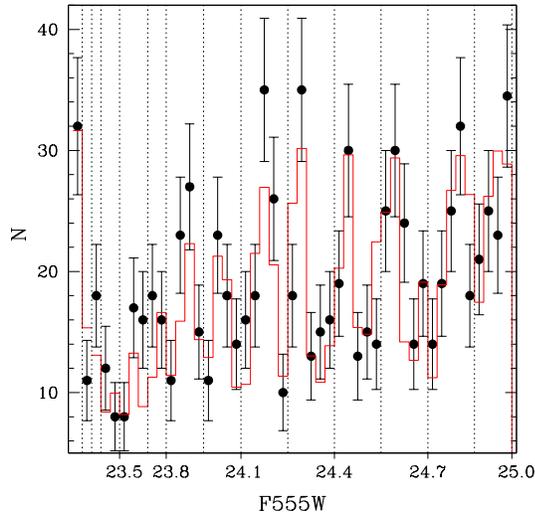}
\figcaption[Best.ps]{The points represent the observed counts in each bin of the
grid shown in Figure~\ref{ACSgrid}. The points are ordered by colour from blue to red,
within a given magnitude bin, and the magnitude increases from left to right. The histogram (shown in red) is the best fit model described in the
text. 
\label{Best}}
\end{figure}

 The comparison between models and data is performed by binning on the grid shown in Figure~\ref{ACSgrid}.
The brightest stars are assigned to a single large bin in order to lessen the sensitivity of our results to
uncertainties in initial parameters. The principal goal of this comparison is to locate the magnitude at which
the cooling rate changes due to the transition to photon-dominated cooling from other processes, be it via neutrinos
or more exotic processes. The colour binning is chosen to constrain the location of the cooling sequence in colour
and thereby measure the extinction simultaneously with the cooling. The final grid thus
 contains 47 bins, so that  comparison with a particular core model 
 has 40 degrees of freedom. The best fit for a model with our
standard neutrino cooling yields $\chi^2 = 44.8$, for $0.53 M_{\odot}$, a Helium
layer mass fraction $q_{He} = 0.037$ and $q_H = 4 \times 10^{-4}$. This Hydrogen
layer is slightly larger than usually assumed, but not so large that residual
Hydrogen burning is a major contributor. Larger masses ($q_H >  6 \times 10^{-4}$)
yield poor fits ($\chi^2 > 100$) because the nuclear contribution dramatically
changes the shape of the cooling curve,  indicating that the cooling of the white dwarfs in 47~Tuc are
not being substantially delayed by residual H burning, as has been suggested for lower metallicity
progenitors (Miller Bertolami, Althaus \& Garcia-Berro 2013). The best fit distance modulus is $\mu_0 = 13.27$,
which compares well with other distance estimates, although the best fit reddening,
E(B-V)=0.08, is at the red end of the allowed range. Figure~\ref{Best} shows the comparison
of the model with the data. In this plot, we show the counts as a function of colour in each luminosity
bin (separated by vertical dotted lines). For instance, the five bins between the dotted lines
indicating F555W=24.1 and 24.25 show the counts in five colour bins spanning $F435W-F555W=-0.1$ to 0.15.
This projection of the two-dimensional data into a pseudo-luminosity function allows us to
assess the quality of the model fit to the data.

\begin{figure}
\plotone{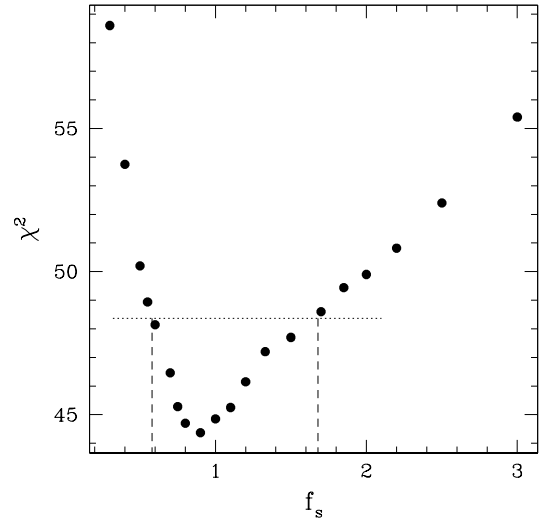}
\figcaption[fChi.ps]{The solid points show the minimum $\chi^2$ at each value of $f_s$, after marginalising
over white dwarf masses and values of $q_{He}$ and $q_H$, as well as distance and extinction. The dotted line
shows the value of $\Delta \chi^2=4$ relative to the minimum, so that the dashed vertical lines indicate the
region of $95\%$ confidence. The slope is steeper for $f_s<1$ than for $f_s>1$, suggesting that the constraints
on a modicum of extra cooling are weaker than models with less cooling.
\label{fChi}}
\end{figure}

We test the sensitivity of our results to the neutrino emission by applying a uniform
scaling factor $f_{s}$ to the neutrino emission and redoing the fit, marginalising over
the white dwarf mass, internal composition, and layer masses, as well as distance modulus and extinction ranges described above.
The resulting curve of $\chi^2$ as a function of $f_s$ is shown in Figure~\ref{fChi}.
 At 95\% confidence, $0.6 < f_s < 1.7$, with the
best fit $\chi^2$ achieved for $f_s \sim 0.9$. The value at $f_s=1$ is only $\Delta \chi^2=0.5$ larger than
the minimum, so that the difference is not statistically significant. Thus, our white dwarf cooling models favour
the traditional implementation of neutrino emission and constrain variations to be
approximately within a factor of two. There is also no indication that increased neutrino cooling can
be offset by nuclear burning in a thick H envelope.

\begin{figure}
\plotone{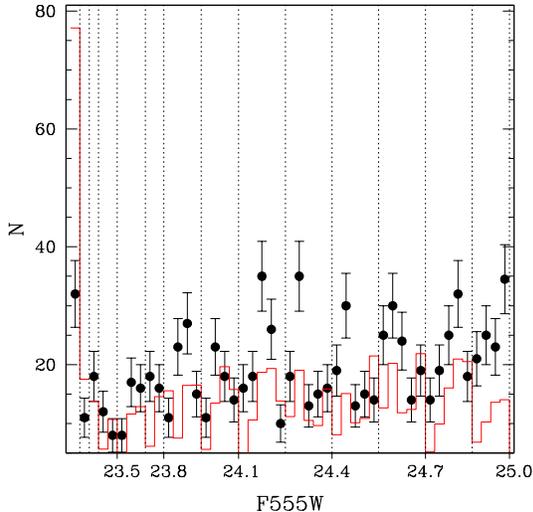}
\figcaption[BestNoNeut.ps]{The points represent the same observed counts as in
Figure~\ref{Best}, but the red histogram is now the best fit model for the
case of no neutrino cooling. We see that the requirement that the model fit the
data at the faint end results in a large overestimate of the number of bright 
white dwarfs.
\label{NoNeut}}
\end{figure}

The $f_s$ scaling models all start with the initial conditions determined from the MESA models, but we noted in
\S~\ref{Model} that a self-consistent evolutionary model with no neutrinos would begin with
a higher central temperature, so we have also tested a model with no neutrino cooling
 starting with the appropriate initial conditions. The best fit is achieved
for a $0.55 M_{\odot}$ model with $q_{He}=0.025$ and $q_H = 2 \times 10^{-4}$. The
best fit distance and extinction is $\mu_0=13.16$ and E(B-V)=0.08. Nevertheless,
the minimum $\chi^2 = 184$, which is very poor compared to our best fit model.
Figure~\ref{NoNeut} shows this model and demonstrates, when compared with Figure~\ref{Best},
how this data discriminates between models. In this case, the lack of neutrino cooling
results in slower early cooling and an excess of counts at bright magnitudes relative to
faint magnitudes. 

These results also allow us to anticipate a comparison with constraints on neutrino emission from the
pulsational stability of hot DBV white dwarfs (Winget et al. 2004). Bischoff-Kim (2008) provides
a model for the pulsational properties of the hot DBV star EC20058-5234, and predicts the
rate of change of two pulsational periods as a function of $\lambda$ (scaled explicitly relative to the
plasmon neutrino rate in that case, but effectively the same as our overall scaling parameter $f_s$ in this regime).
Based on our constraints above and the figures in Bischoff-Kim (2008), we anticipate that the
period derivatives of the 257s and 281s modes in that star should be within $1.0\pm 0.4 \times 10^{-13} s/s$
and $ 1.4 \pm 0.5 \times 10^{-13} s/s$ respectively.

\subsection{The Neutrino Magnetic Moment}

The restriction on scaled neutrino emissivities should then also translate into constraints
on exotic radiation mechanisms. The constraint $f_s<1.7$ can be used in equation~(\ref{Eps_q})
to obtain an approximate limit $\mu_{\nu} < 5 \times 10^{-12} \mu_B$. A more accurate limit
is obtained by repeating the cooling calculations using the correction to the emissivity
given by equation~(\ref{Eps_q}), including the density dependance. If we repeat the fitting
procedure with these models, marginalising again over distance, extinction and white dwarf mass parameters, we find that the 95\% confidence limit on the neutrino magnetic moment is
$\mu_{\nu} < 3.4 \times 10^{-12} \mu_B$.

This limit is an order of magnitude better than the current laboratory limits $\sim 3 \times 10^{-11} \mu_B$ (Daraktcheva et al. 2005; Arpesella et al. 2008; Beda et al. 2010)
 and also better
than astrophysical limits based on the lifetime of the Sun (Bernstein et al. 1963; Raffelt 1999)
and Horizontal Branch Stars (Raffelt et al. 1989). Most directly, it represents a factor of several improvement
of earlier determinations $\mu_{12}<10$ based on the hot white dwarf luminosity function (Blinnikov \& Dunina-Barkovskaya 1994).
Although measurements of the hot white dwarf luminosity function have improved since then, the excess counts at the bright
end (see discussion in \S~\ref{Bolo}) have hampered the statistical significance of such constraints using the more up-to-date data
 (Miller Bertolami 2014).
Our measurement is also comparable to the limits
derived from the core mass of red giant stars (Castellani \& degl'Innocenti 1993; Catelan et al. 1996;
Haft et al. 1994; Raffelt 1990; Raffelt \& Weiss 1992; Viaux et al. 2013) but has the advantage of being measured
in stars whose properties are dominated by the neutrino emission and in a manner in which the distance (the
largest source of error in the red giant measurements) is constrained simultaneously.

In addition to the consequences for particle physics, a neutrino magnetic moment in
the range $\mu_{12} \sim$ 20--50 could lead to qualitatively different late-time evolution
for stars in the initial mass range $\sim$9--17$M_{\odot}$ (Heger et al. 2009). However, our
limit here appears to preclude this eventuality.

\subsection{The Axion Mass}

In a similar fashion, models with additional cooling due to axion emission, given by equation~(\ref{Eps_ax}), can be compared
to the data to obtain a 95\% confidence  limit of $g_{ae} < 8.4 \times 10^{-14}$, which implies an axion mass $< 4.1 meV$.

This constraint contradicts the claim of Corsico et al. (2012), whose modelling of the pulsating white dwarf G117-B15.A implied that
the default rate of cooling was too slow to match the observed rate of change of the pulsation period. They invoked an extra cooling
due to axion emission that implied an axion mass of $m_{ax} = 17.4^{+2.3}_{-2.7} meV$, a 5.8 $\sigma$ discrepancy with respect to
our limit. However, the Corsico result was based on the rate of change of a particular oscillation mode which they claimed was
trapped in the Hydrogen envelope. The trapping strongly slows the expected rate of period change and suggests that future improvements
in the asteroseismological modelling with untrapped modes may better match the observations.

\begin{figure}
\plotone{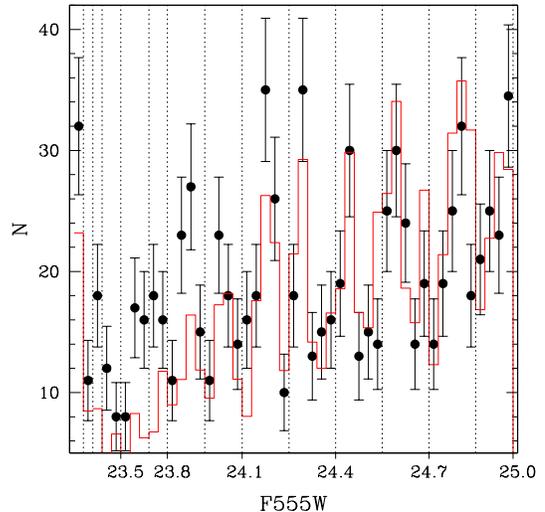}
\figcaption[BestAx.ps]{The points represent the same observed counts as in
Figure~\ref{Best}, but the red histogram is now the best fit model for the
case of axion cooling with an axion mass of 17~meV. Fitting the model at the
faint end results in an underproduction of stars at brighter magnitudes.
\label{Corsico}}
\end{figure}

Figure~\ref{Corsico} shows the best fit solution for the case of a 17~meV neutrino (with a distance $\mu_0=13.21$ and E(B-V)=0.08).
A comparison with Figure~\ref{Best} demonstrates how too much cooling adversely affects the model fit -- namely that the counts
at brighter magnitudes are now too low with respect to fainter magnitudes (the opposite trend seen in Figure~\ref{NoNeut}).

Our limits are  not consistent at the 95\% confidence level with the proposal that an axion mass $\sim 5 meV$ is motivated by the overall behaviour of the hot
white dwarf luminosity function (Isern et al. 2008). Caution has been expressed about this result by Miller Bertolami
et al. (2014), who note that the strength of this conclusion varies depending on which observational luminosity function
is used, and is further muddied by the aforementioned uncertainties in the bright luminosity function. Our results do not require such an axion, as
the inclusion of axion cooling only worsens the fit of the model, but are not strong enough to conclusively rule it out.

\subsection{Radiation into Extra Dimensions}

Testing models which include radiation into extra dimensions place a 95\% limit on the mass scale $M_s > 0.54 TeV/c^2$ when
we include our ad-hoc corrections for degeneracy (ignoring the degeneracy correction increases this to $0.9 TeV/c^2$).
The limits are different  because
 the steep temperature dependance of $\epsilon_{\gamma \gamma}$
means that this rate is irrelevant even for our hot white dwarfs, and only becomes important at
energies corresponding to supernova cores, so that the cooling of the white dwarfs is indeed regulated by the
processes that depend on electron scattering.
 These expressions are comparable to constraints based
on the properties of the Sun but weaker than those based on red giants and
supernovae (Barger et al. 1999, Cassisi et al. 2000). They are also weaker than the constraint derived
by Biesiada \& Malec (2002), although that estimate is likely overstated because the rates were not
corrected for the effects of degeneracy.

\section{Conclusion}

Our results suggest that the cooling of hot white dwarfs is presently well understood. The bolometric luminosity
function we infer from our hottest stars shows a monotonic decrease to the highest effective temperatures observed,
and does not display the plateau observed in some prior studies. This supports recent claims that this feature was
the result of systematic uncertainties rather than the consequence of unanticipated physical effects.

However, our observations in the core of the cluster result in a problematic match between the data and the models, in
the sense that the model colours appear too blue relative to the data at fainter magnitudes. This suggests that either
the photometric calibrations are systematically biased by the crowding of the field or that the
short wavelength opacities are not fully understood for white dwarfs of effective temperatures $\leq$~19,000~K.

The atmosphere models do fit the data for the outer cluster fields, observed in longer wavelength bandpasses.
We fit detailed cooling models to the number counts in these  fields, which provide an excellent match to
the observations without requiring any substantial modifications to the physics. Indeed, the comparison suggests 
that the neutrino emission models of Itoh et al. (1996) are accurate to within a factor of two (at 95\% confidence).
This match can also be used to constrain a variety of proposals for additional exotic cooling mechanisms, and provides
constraints on the existence of a neutrino magnetic moment, emission of axions, and possible radiation of energy into
extra dimensions. In each of these cases, the inclusion of extra cooling makes the fit to the data worse, and provides
little evidence for any exotic physics.

\acknowledgements 

This research is based on NASA/ESA Hubble Space Telescope observations obtained at the Space Telescope Science Institute, which is operated by the Association of Universities for Research in Astronomy Inc. under NASA contract NAS5-26555. These observations are associated with proposal GO-12971 (PI: Richer). This work was supported by NASA/HST grants GO-12971, NSF grant AST-1211719, the Natural Sciences and Engineering Research Council of Canada, the Canadian Foundation for Innovation, the British Columbia Knowledge Development Fund.
 It has made use of the NASA ADS and arXiv.org databases.

\end{document}